\documentclass[%
reprint,
superscriptaddress,
 amsmath,amssymb,
 aps,
prl,
nobibnotes,]{revtex4-2}

\usepackage{graphicx}
\usepackage{dcolumn}
\usepackage{bm}
\usepackage{xcolor}
\usepackage{multirow}

\newcommand*{\MIT }{Massachusetts Institute of Technology, Cambridge, Massachusetts 02139, USA}
\newcommand*{\ODU}{Old Dominion University, Norfolk, Virginia 23529, USA}
\newcommand*{\JLAB}{Thomas Jefferson National Accelerator Facility, Newport News, Virginia 23606, USA}
\newcommand*{\TAU }{School of Physics and Astronomy, Tel Aviv University, Tel Aviv 69978, Israel}

\newcommand*{\GW}{The George Washington University, Washington, D.C., 20052, USA}

\newcommand*{\Duke}{Duke University and Triangle Universities Nuclear Laboratory, Durham, NC 27708, USA}
\newcommand*{\UNH}{University of New Hampshire, Durham, New Hampshire 03824, USA}
\newcommand*{\MSU}{Mississippi State University, Mississippi State, MS 39762, USA}
\newcommand*{\FIU}{Florida International University, Miami, FL 33199, USA}

\newcommand*{\CMU}{Carnegie Mellon University, Pittsburgh, PA 15213, USA}

\newcommand*{\FSU}{Florida State University, Tallahassee, FL 32306, USA}

\newcommand*{\UNCW}{University of North Carolina Wilmington, Wilmington, NC 28403, USA}

\newcommand*{\VT}{Virginia Tech, Blacksburg, VA 24061, USA}

\newcommand*{\Lamar}{Lamar University, Beaumont, TX 77705, USA}

\newcommand*{\WM}{College of William and Mary, Williamsburg, VA 23185, USA}

\newcommand*{\NCATSU}{North Carolina Agricultural and Technical State University, Greensboro, NC 27411, USA}

\newcommand*{\ASU}{Arizona State University, Tempe, AZ 85287, USA}

\newcommand*{\Glascow}{University of Glasgow, Glasgow G12 8QQ, Scotland, UK}

\newcommand*{\UCONN}{University of Connecticut, Storrs, CT 06269, USA}

\newcommand*{\Yerevan}{A. I. Alikhanyan National Science Laboratory (Yerevan Physics Institute), 0036, Yerevan, Armenia}

\newcommand*{\SBU}{Stony Brook University, Stony Brook, NY 11794, USA}

\newcommand*{\TOMSK}{Tomsk State University, 634050 Tomsk, Russia}

\newcommand*{\WJ}{Washington and Jefferson College, Washington, PA 15301, USA}

\newcommand*{\OSAKA}{Osaka University, Osaka, Japan}

\newcommand*{\REGINA}{University of Regina, Regina, SK S4S 0A2, Canada}

\newcommand*{\GSID}{GSI Helmholtz Centre for Heavy Ion Research, Darmstadt, Germany}

\newcommand*{\RUB}{Ruhr University Bochum, Bochum, Germany}

\newcommand*{\UMASSA}{University of Massachusetts Amherst, Amherst, MA 010031, USA}

\newcommand*{\NRNUM}{National Research Nuclear University MEPhI, Moscow 115409, Russia}

\newcommand*{\NRCKI}{National Research Centre Kurchatov Institute, Moscow 123182, Russia}

\newcommand*{\UNT}{University of North Texas, Denton, TX 76201, USA}

\newcommand*{\Wuhan}{Wuhan University, Wuhan, Hubei, China}

\newcommand*{\UTK}{The University of Tennessee, Knoxville, TN 37996, USA}

\begin{document}
\setlength{\extrarowheight}{2pt}


\title{First Measurement of Near- and Sub-Threshold $J/\psi$  Photoproduction off Nuclei} 

\author{J.R.~Pybus}
\email{jrpybus@mit.edu}
\affiliation{\MIT}

\author{L.~Ehinger}%
\affiliation{\MIT}%

\author{T.~Kolar}
\affiliation{\TAU}

\author{B.~Devkota}
\affiliation{\MSU}
\author{P.~Sharp}
\affiliation{\GW}
\author{B.~Yu}
\affiliation{\Duke}

\author{M.M.~Dalton}
\affiliation{\JLAB}
\author{D.~Dutta}
\affiliation{\MSU}
\author{H.~Gao}
\affiliation{\Duke}
\author{O.~Hen}
\affiliation{\MIT}
\author{E.~Piasetzky}
\affiliation{\TAU}
\author{S.N.~Santiesteban}
\affiliation{\UNH}
\author{A.~Schmidt}
\affiliation{\GW}
\author{A.~Somov}
\affiliation{\JLAB}
\author{H.~Szumila-Vance}
\affiliation{\JLAB}

\author{S.~Adhikari}
\affiliation{\ODU}
\author{A.~Asaturyan}
\affiliation{\JLAB}
\author{A.~Austregesilo}
\affiliation{\JLAB}
\author{C.~Ayerbe Gayoso}%
\affiliation{\MSU}
\author{J.~Barlow}
\affiliation{\FSU}
\author{V.V.~Berdnikov}
\affiliation{\JLAB}
\author{H.D.~Bhatt}
\affiliation{\MSU}
\author{Deepak Bhetuwal}
\affiliation{\MSU}
\author{T.~Black}
\affiliation{\UNCW}
\author{W.J. ~Briscoe}
\affiliation{\GW}
\author{G.~Chung}
\affiliation{\VT}
\author{P.L.~Cole}
\affiliation{\Lamar}
\author{A.~Deur}
\affiliation{\JLAB}
\author{R.~Dotel}
\affiliation{\FIU}
\author{H.~Egiyan}
\affiliation{\JLAB}
\author{P.~Eugenio}
\affiliation{\FSU}
\author{C.~Fanelli}
\affiliation{\WM}
\author{L.~Gan}
\affiliation{\UNCW}
\author{A.~Gasparian}
\affiliation{\NCATSU}
\author{J.~Guo}
\affiliation{\CMU}
\author{K.~Hernandez}
\affiliation{\ASU}
\author{D.W.~Higinbotham}
\affiliation{\JLAB}
\author{P.~Hurck}
\affiliation{\Glascow}
\author{I.~Jaegle}
\affiliation{\JLAB}
\author{R.T.~Jones}
\affiliation{\UCONN}
\author{V.~Kakoyan}
\affiliation{\Yerevan}
\author{A.~Karki}
\affiliation{\MSU}
\author{H.~Li}
\affiliation{\WM}
\author{W.B.~Li}
\affiliation{\SBU}
\author{G.R.~Linera}
\affiliation{\FSU}
\author{V.~Lyubovitskij}
\affiliation{\TOMSK}
\author{H.~Marukyan}
\affiliation{\Yerevan}
\author{M.D.~McCaughan}
\affiliation{\JLAB}
\author{M.~McCracken}
\affiliation{\WJ}
\author{K.~Mizutani}
\affiliation{\OSAKA}
\author{D.~Nguyen}
\affiliation{\UTK}
\author{S.~Oresic}
\affiliation{\REGINA}
\author{A.I.~Ostrovidov}
\affiliation{\FSU}
\author{Z.~Papandreou}
\affiliation{\REGINA}
\author{C.~Paudel}
\affiliation{\FIU}
\author{K.~Peters}
\affiliation{\GSID}
\author{J.~Ritman}
\affiliation{\GSID}
\affiliation{\RUB}
\author{A.~Schick}
\affiliation{\UMASSA}
\author{J.~Schwiening}
\affiliation{\GSID}
\author{A.~Smith}
\affiliation{\JLAB}
\author{S.~Somov}
\affiliation{\NRNUM}
\author{I.~Strakovsky}
\affiliation{\GW}
\author{K.~Suresh}
\affiliation{\REGINA}
\author{V.V.~Tarasov}
\affiliation{\NRCKI}
\author{S.~Taylor}
\affiliation{\JLAB}
\author{T.~Xiao}
\affiliation{\UNT}
\author{Z.~Zhang}
\affiliation{\Wuhan}
\author{X.~Zhou}
\affiliation{\Wuhan}


\date{\today}

\begin{abstract}

We report on the first measurement of $J/\psi$ photoproduction from nuclei in the photon energy range of $7$ to $10.8$ GeV, extending above and below the photoproduction threshold in the free proton of $\sim8.2$ GeV.
The experiment used a tagged photon beam incident on deuterium, helium, and carbon, and the GlueX detector at Jefferson Lab to measure the semi-inclusive $A(\gamma,e^+e^-p)$ reaction with a dilepton invariant mass $M(e^+e^-)\sim m_{J/\psi}=3.1$ GeV.
The incoherent $J/\psi$ photoproduction cross sections in the measured nuclei are extracted as a function of the incident photon energy, momentum transfer, and proton reconstructed missing light-cone momentum fraction.
Comparisons with theoretical predictions assuming a dipole form factor allow extracting a gluonic radius for bound protons of $\sqrt{\langle r^2\rangle}=0.85\pm0.14$ fm. The data also suggest an excess of the measured cross section for sub-threshold production and for interactions with high missing light-cone momentum fraction protons. The measured enhancement can be explained by modified gluon structure for high-virtuality bound-protons.

\end{abstract}

\maketitle


Understanding the nuclear-medium modification of the partonic structure of nucleons has been a perennial problem in nuclear physics~\cite{Arnold:1984,Aubert83,Ashman88,Gomez94,Arneodo90,Seely09,Schmookler:2019nvf,Hen:2016kwk}.
Despite tremendous advances in modeling existing data, the underlying mechanisms of the nuclear modification of valence quark distributions remains elusive~\cite{Frankfurt88,Sargsian02,Norton03,Hen:2016kwk}.
The problem is further complicated by indications that the gluon distributions are also modified in nuclei~\cite{annurev:/content/journals/10.1146/annurev-nucl-011720-042725,ncteqcollaboration2023evidencemodifiedquarkgluondistributions}.

Measurements of the photoproduction of $J/\psi$ mesons is a well-known tool for studying gluon density and its spatial distributions in nucleons. 
Recent studies of $J/\psi$ photoproduction off the proton in the near-threshold photon energy region provided new insights into the gluonic content of the proton at large momentum fraction~\cite{PhysRevLett.123.072001,GravitationalFF,PhysRevC.108.025201}.
Similarly, measurements of $J/\psi$ photoproduction from nuclei has the potential to provide unique insights into the gluonic content of nuclei and bound nucleons, though the specific reaction mechanisms contributing to near-threshold photoproduction of $J/\psi$ are currently under theoretical debate~\cite{PhysRevC.108.025201}.
In particular, the ``sub-threshold'' photoproduction of $J/\psi$, using photons with energy below $E_\gamma^{th}\approx8.2$ GeV, has long been sought after as a signature of high-energy gluon configurations in the nucleus~\cite{HOYER1997c284,hoyer1997charmonium,PhysRevC.79.015209,PhysRevC.109.065206}.

Recent studies suggest that high-momentum nucleons in Short-Range Correlated (SRC) pairs configurations~\cite{subedi08,hen14,Hen:2016kwk} may be the dominant source of gluons with large momentum-fraction $x$~\cite{Xu_2020,Hatta_2020}. Therefore, measuring the sub-threshold $J/\psi$ photoproduction process could be sensitive to modification of the gluons in SRC nucleons and other exotic effects such as hidden-color components of the nucleus~\cite{Brodsky_2001}.

Data on near and sub-threshold $J/\psi$ photoproduction off nuclei is sparse. Early measurements at low-energies~\cite{PhysRevLett.35.1616,PhysRevLett.35.483} did not tag the beam photons and therefore cannot differentiate as a function of photon energy. A more recent dedicated search for sub-threshold $J/\psi$ production was made at energies far below threshold and observed no $J/\psi$ events~\cite{PhysRevC.79.015209}. 





Here we present the first measurement of $J/\psi$ production from nuclear targets in the near- and sub-threshold region of $7<E_\gamma<10.8$ GeV using a tagged photon beam via the semi-inclusive $A(\gamma,e^+e^-p)$, following the leptonic decay $J/\psi\rightarrow e^+e^-$.
The detection of a coincident proton allows both an improved reconstruction of the dilepton invariant mass and an inference of the initial-state nucleon momentum, providing further insight to the nuclear effects present in the reaction.
We extract differential cross sections for the measured reaction and compare with theoretical calculations.


The experiment ran in 2021 in Hall D of the Thomas Jefferson National Accelerator Facility. 
A 10.8 GeV electron beam from the Continuous Electron Beam Accelerator Facility~\cite{doi:10.1146/annurev.nucl.51.101701.132327} was used to create a tagged linearly-polarized photon beam via coherent bremsstrahlung from a diamond radiator~\cite{GlueXNIM}. 
The energy, $E_\gamma$, of the bremsstrahlung photons follows a spectrum proportional to $1/E_\gamma$, with coherent enhancements primarily occurring at an energy 7.6-8.6 GeV.
The photon energy was determined from the measured momentum of the scattered electron, determined by a tagging Microscope and Hodoscope to an accuracy of about 0.1\%.
The photon beam was collimated 75 meters downstream of the radiator, before reaching the the target within the GlueX spectrometer.
The photon flux incident on the target was monitored by the Pair Spectrometer (PS)~\cite{PS}, allowing for a measurement of the energy-dependent luminosity.

Two 30-cm-long liquid targets of $^2$H and $^4$He, with total tagged photon-on-nucleon luminosity of 33 and 60 pb$^{-1}$ in the region $E_\gamma>7$ GeV, respectively, were used. 
A solid multifoil $^{12}$C target (8 equidistant foils with a total thickness of 1.9 cm, extended over a 30-cm region) was also used, with a total tagged photon-on-nucleon luminosity of 94 pb$^{-1}$ in the region $E_\gamma>7$ GeV.

Outgoing particles were measured using the large-acceptance GlueX spectrometer~\cite{GlueXNIM}. GlueX is composed of multiple subdetectors. 
Immediately surrounding the target is a scintillator-based start counter (SC)~\cite{SC}, followed by a straw-tube central drift chamber (CDC)~\cite{CDC}, and lead and scintillating-fiber barrel calorimeter (BCAL)~\cite{BCAL}, encompassed within a superconducting solenoid magnet. 
Further downstream in the incident beam direction are a set of planar forward drift chambers (FDC)~\cite{FDC}, a time-of-flight scintillator detector (TOF), and a lead-glass forward calorimeter (FCAL)~\cite{FCAL}. 
Events were recorded when sufficiently large energy was deposited in the calorimeters or a small energy deposition was paired with a hit in the start counter.

We study the semi-inclusive reaction $\gamma A\rightarrow e^+e^-p(X)$, where $X$ denotes the undetected residual nuclear state.
The charged particles were measured by the drift chambers, with the curvature of their tracks in the magnetic field used to determine their momentum.
Tracking resolution was improved using a kinematic fit constraining a common reaction vertex in the target for all measured tracks.

The identification of electrons and rejection of pions was performed using two calorimetry methods, following Ref.~\cite{PhysRevLett.123.072001,PhysRevC.108.025201}.
First, selections were applied on the momentum-energy ratio $p/E$, where the charged particle momentum $p$ was determined using the charged track information from the drift chambers, and the energy $E$ was determined from the energy deposition in the calorimeters. 
Because electrons and positrons deposit the majority of their energy in the calorimeters, in contrast to heavier particles, this value is expected to be close to 1.
As in previous studies, this ratio was constrained to $-3\sigma<p/E - \langle p/E \rangle < 2\sigma$, where $\sigma$ is the resolution of $p/E$ in each calorimeter, found to be 8\% for the FCAL and 7\% for the BCAL.
The resolution for FCAL candidates is roughly twice that of previous GlueX measurements, where exclusivity constraints on reactions from the hydrogen target enabled improved momentum resolution for forward-going particles; for BCAL candidates, the resolution is similar.

The second method of $e/\pi$ separation used the innermost layer of the BCAL as a preshower detector, requiring the energy $E_{pre}$ deposited in this layer by a lepton candidate to satisfy $E_{pre}\sin \theta>30$ MeV, where the sine of the charged particle angle $\theta$ accounts for the path length of the particle in the BCAL layer. 
As electrons and positrons deposit much more energy in this layer than pions, this selection cut also rejects a large fraction of the pion backgrounds. 

Proton identification was performed by comparing the track momentum with time-of-flight signals provided by either the FCAL, BCAL, or TOF. Additional discrimination was achieved by considering energy loss energy loss in the CDC, FDC, and SC.

Tagged beam photons were associated with an event if they fell within a 2 ns synchronous with the arrival of the photon at the target.
The substantial background caused by accidental coincident tagged photons was estimated and subtracted by using the rate of ``off-time photons'' with tagger signals between 6 and 18 ns before or after the synchronous time.

Additional cuts were applied on the events to improve signal-to-background ratios. 
All charged particles were required to have a momentum $>0.4$ GeV$/c$, as well as a polar angle $>2^\circ$, in order to stay within the fiducial region of the detector where the acceptance and response are well-understood. 
The vertex of the event was required to fall within the volume of the target. 
An ``elasticity'' requirement was also placed on the events, requiring energy balance within 1 GeV between the initial- and final-state particles assuming scattering from a quasi-free proton.
Finally, events with extra tracks in the drift chambers or showers in the calorimeters were removed.

The $J/\psi$ yield was determined by examining the invariant mass spectrum of the final-state dilepton. 
The poor dilepton invariant mass reconstruction resolution substantially reduced our ability to isolate $J/\psi$ signal over background, resulting from residual $\pi^+\pi^-$ pion misidentification and Bethe-Heitler $e^+e^-$ production.

To improve the mass resolution, we make the observation that, along with the transverse momentum $p_\perp$, the light-cone coordinate ``minus'' component of momentum $p^-\equiv E-p_z$ is well reconstructed for the high-momentum leptons, where $E$ and $p_z$ are the energy and longitudinal component of momentum relative to the beamline. 
Furthermore, the low-momentum protons and the tagged beam photons are also well-reconstructed; if we assume that the initial momentum of the struck nucleon is balanced by a single final-state partner nucleon, the quasi-elastic $J/\psi$ production can be approximated as originating from a pair of nucleons at rest.
Under this approximation, we calculate the dilepton mass using only well-measured quantities as:
\begin{align}
\begin{split}
    m^2(e^+e^-)\approx m_\text{light-cone}^2\hspace{15 em} \\
    =\left(p_{e^+}^-+p_{e^-}^-\right)
    \left(2E_\gamma + 2m_N -p_p^+ - \frac{m_{N}^2+p_\text{tot}^2}{2m_N -p_\text{tot}^-}\right) \\
    - \left(\vec p_{e^+}^\perp+\vec p_{e^-}^\perp\right)^2\,.
\end{split}
\end{align}
where $m_N$ is the nucleon mass, $p^+\equiv E + p_z$ is the ``plus'' component of momentum, and $p_\text{tot}=p_{e^+} + p_{e^-} + p_{p}$ is the total four-momentum of the measured final-state.

This ``light-cone'' mass proxy allows substantially improved resolution and the isolation of $J/\psi$ events above background. 
The resolution and accuracy of the light-cone mass has been studied via full simulations of the $\gamma A\rightarrow J/\psi p X$ reactions in the GlueX spectrometer, and found to have improved resolution while being robust to the impacts of nuclear motion and proton rescattering. 
Further details of the derivation of the light-cone mass, as well as simulation studies of its behavior, are given in the Supplemental Materials.

Figure~\ref{fig:insert_mass} shows the measured $m_\text{light-cone}$ distribution for the combined data of all targets. 
The peak resulting from the $J/\psi\rightarrow e^+e^-$ decay can be clearly seen.
The inset shows the same for events in the ``sub-threshold'' region $E_\gamma<8.2$ GeV where we also observe the $J/\psi\rightarrow e^+e^-$ decay peak resulting from sub-threshold production of $J/\psi$. 
The statistical significance of the sub-threshold peak, when compared to the background-only fit to data, is found to be $3.25\sigma$ when allowing the signal parameters to vary; when fixing the signal shape to that of the above-threshold mass peak, the significance is slightly reduced to $2.86\sigma$.

\begin{figure}[t]
    \centering
    \includegraphics[width = 0.45 \textwidth]{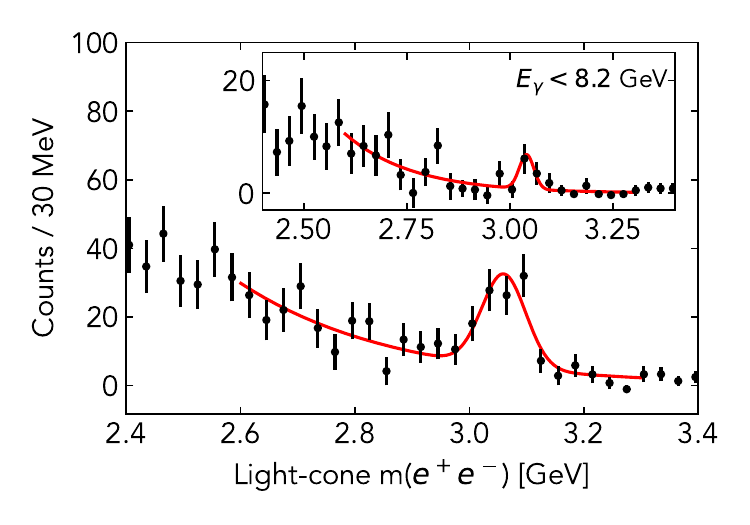}
    \caption{\emph{Main figure:} Light-cone mass distribution for the combined data of helium and carbon, fit using an exponential background and a Gaussian signal. \emph{Inset:} Light-cone mass distribution for events in the ``sub-threshold'' region $E_\gamma<8.2$ GeV. In both cases the $J/\psi$ decay can be clearly seen.}
    \label{fig:insert_mass}
\end{figure}


We calculate the total cross section as a function of the incoming photon energy as:
\begin{equation}
    \sigma(E_\gamma)=\frac{Y_{J/\psi}(E_\gamma)}{{\mathcal L}(E_\gamma)\epsilon(E_\gamma)T_A{\mathcal B}(J/\psi\rightarrow e^+e^-)}\,,
\end{equation}
where $Y_{J/\psi}$ is the extracted yield of $J/\psi \rightarrow e^+e^-$ decays, ${\mathcal L}$ is the tagged photon luminosity, $\epsilon$ is the reconstruction efficiency as determined from Monte-Carlo simulations, $T_A$ is the proton transparency for nucleus $A$~\cite{abbott98,osti_785455,DEVRIES1987495,osti_785455,PhysRevC.45.791} (detailed in the Supplemental Material), and ${\mathcal B}(J/\psi\rightarrow e^+e^-)\approx5.97\pm0.03$\% is the branching fraction of $J/\psi$ into $e^+e^-$~\cite{PDG}. 

For the total cross section extraction, $Y_{J/\psi}$ is determined by performing an unbinned likelihood fit on the light-cone mass distribution, assuming an exponential background.
For the differential cross sections, it is determined by performing side-band subtraction using the light-cone mass distribution region of $2.7<m_\text{light-cone}<3.4$ GeV, as a measure of the background contribution, excluding the signal region of $\pm3\sigma$ around the $J/\psi\rightarrow e^+e^-$ peak.

The dominant point-to-point systematic uncertainty is due to the event selection cuts, determined by varying the values of the lepton PID and energy balance cuts and then taking the resulting variance on the extracted cross section, accounting for both the change in measured yield and simulated reconstruction efficiency.
We also considered the uncertainty in the energy-dependence of the measured luminosity resulting from the acceptance of the PS, uncertainty in the efficiency as a function of final-state kinematics, and uncertainty on the method used to extract the $J/\psi$ yield.

Normalization uncertainty was determined for each nucleus and ranges from 20\% to 24\%.
This uncertainty is dominated by the uncertainty on the Monte-Carlo efficiency calculations; Ref.~\cite{PhysRevC.108.025201} estimated a 19.5\% uncertainty using $e^+e^-$ production from the Bethe-Heitler process and simulation to benchmark the exclusive measurement $\gamma p\rightarrow J/\psi p$, and we refer to this estimate.
Additional sources of normalization uncertainty include the proton transparency and the target thickness and density.

\begin{table}[b]
\caption{Total per-proton cross sections for A$(\gamma,J/\psi p)X$ luminosity-averaged over the energy range $7<E_\gamma<10.6$ GeV. Data also includes a common 20\% normalization uncertainty (not shown); other normalization uncertainties differ across nuclei and are incorporated into the total cross sections here.}
\label{tab:cs_A}
\begin{tabular}{|l|l|l|l|l|}
\hline
Nucleus   & Plane-wave  & Measured & Statistical    & Systematic     \\
    & cross section  & cross section & uncertainty    & uncertainty     \\\hline
$^2$H & 0.24 nb    & 0.23 nb  & 0.07 nb & 0.04 nb \\ \hline
$^4$He    & 0.22 nb    & 0.33 nb  & 0.06 nb & 0.05 nb \\ \hline
$^{12}$C    & 0.24 nb    & 0.25 nb  & 0.05 nb & 0.05 nb \\ \hline
\end{tabular}
\end{table}

Table~\ref{tab:cs_A} lists the total measured and Plane-Wave Impulse Approximation (PWIA) calculated cross section for each nucleus. Both the measurement and calculation are averaged over the photon flux for each target.


The PWIA calculations assumed a factorized cross section model for the quasi-elastic process $(\gamma,J/\;\psi p)$, given by~\cite{Hatta_2020}:
\begin{equation}
    \frac{d\sigma(\gamma A\rightarrow J/\psi\; p \;X)}{dt\; d^3{\vec p}_{i}\; d E_{i}} = v_{\gamma i}\cdot \frac{d\sigma}{dt}(\gamma p\rightarrow J/\psi\; p)\cdot S(p_{i},E_{i})
\end{equation}
where $p_i=\left(E_i,{\vec p}_{i}\right)$ is the 4-momentum of the struck proton $i$ inside the nucleus, $p_\gamma$ is the 4-momentum of the incoming beam photon, $v_{\gamma i}=p_\gamma\cdot p_i/(E_\gamma E_i)$ is the relative velocity between the photon and the struck proton, and the differential cross section $d\sigma/dt$ for the exclusive process $(\gamma p\rightarrow J/\psi p)$ was taken from a fit to GlueX data~\cite{PhysRevC.108.025201}.
The spectral functions $S(p_i,E_i)$ for helium and carbon were provided by Ref.~\cite{VMC:MFSpecFnc} for mean-field (low-momentum) protons and the Generalized Contact Formalism~\cite{Weiss:2018tbu,schmidt20,Pybus:2020itv} for the SRC protons, calculated using the phenomenological AV18 interaction~\cite{veerasamy11}. 
The momentum distribution for deuterium was taken from Ref.~\cite{wiringa14}, again calculated using the AV18 interaction.
The produced $J/\psi$ was assumed to decay to $e^+e^-$ in a helicity-conserving manner.

We performed Monte Carlo simulation of quasi-elastic $J/\psi$ production from $^2$H, $^4$He, and $^{12}$C, in order to compare with distributions in data as well as estimate the event detection efficiency.
Generated events were simulated using the GEANT model of the GlueX detector~\cite{GlueXNIM}, and reconstructed in the same manner as the measured data.
This simulated detector response was superimposed with randomly triggered samples of data, in order to account for photon tagger accidentals and detector pileup.
These simulations were used to extract the reconstruction efficiency for $J/\psi p$ events, found to be $\sim$13\% with little variation as a function of beam photon energy or nucleus. 
Following a recent study of the QED Bethe-Heitler process with GlueX~\cite{PhysRevC.108.025201} that found the experimental $e^+e^-p$ efficiency to be $85\pm2\%$ of that predicted by simulation, we rescaled our simulated efficiency by that factor to arrive at an estimated average efficiency of $\sim$11\%.

\begin{figure}[t]
    \centering
    \includegraphics[width = 0.45 \textwidth]{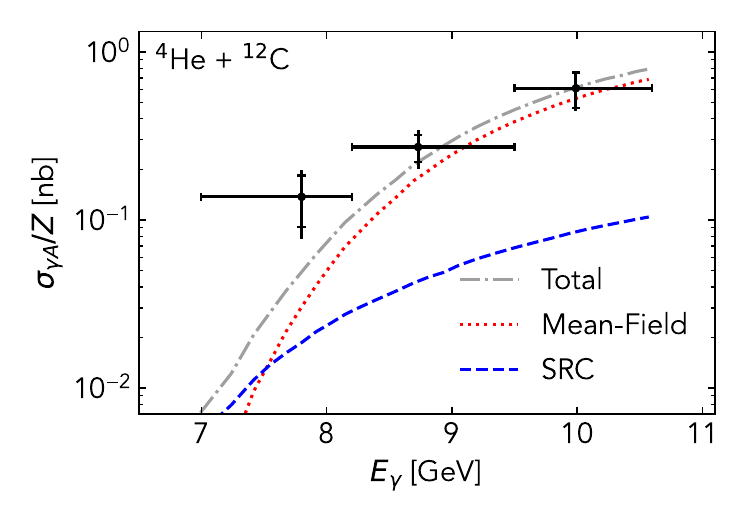}
    \caption{Luminosity-weighted average of the A$(\gamma,J/\psi p)X$ cross section for $^4$He and $^{12}$C, compared with plane-wave calculations for this average. The measured cross section (black) is compared with plane-wave calculations, including the mean-field (dotted red) and SRC (dashed blue) contributions as well as the total (dot-dashed grey). Data also includes a common 23\% normalization uncertainty (not shown).}
    \label{fig:avg_cs}
\end{figure}

Figure~\ref{fig:avg_cs} shows the energy-dependence of the measured cross sections for $^4$He and $^{12}$C, where the result has been combined in a yield-weighted fashion to improve the statistical precision. 
Data are compared with PWIA calculations, separating the contribution from photon interaction with mean-field and short-range correlated nucleons.
We observe that both the $A$-dependence (given in Table~\ref{tab:cs_A}) and the energy-dependence of the measured cross sections are largely consistent with the predictions of plane-wave calculations.
We also find that we are able to measure the cross section in the sub-threshold energy region $E_\gamma<8.2$ GeV, marking the first such measurement of $J/\psi$ production below the proton energy threshold.
The sub-threshold cross section appears to somewhat exceed the plane-wave predictions, indicating the possibility of more exotic mechanisms at play, though higher statistics data are needed to make a more definitive claim.


\begin{figure}[t]
    \centering
    \includegraphics[width = 0.45 \textwidth]{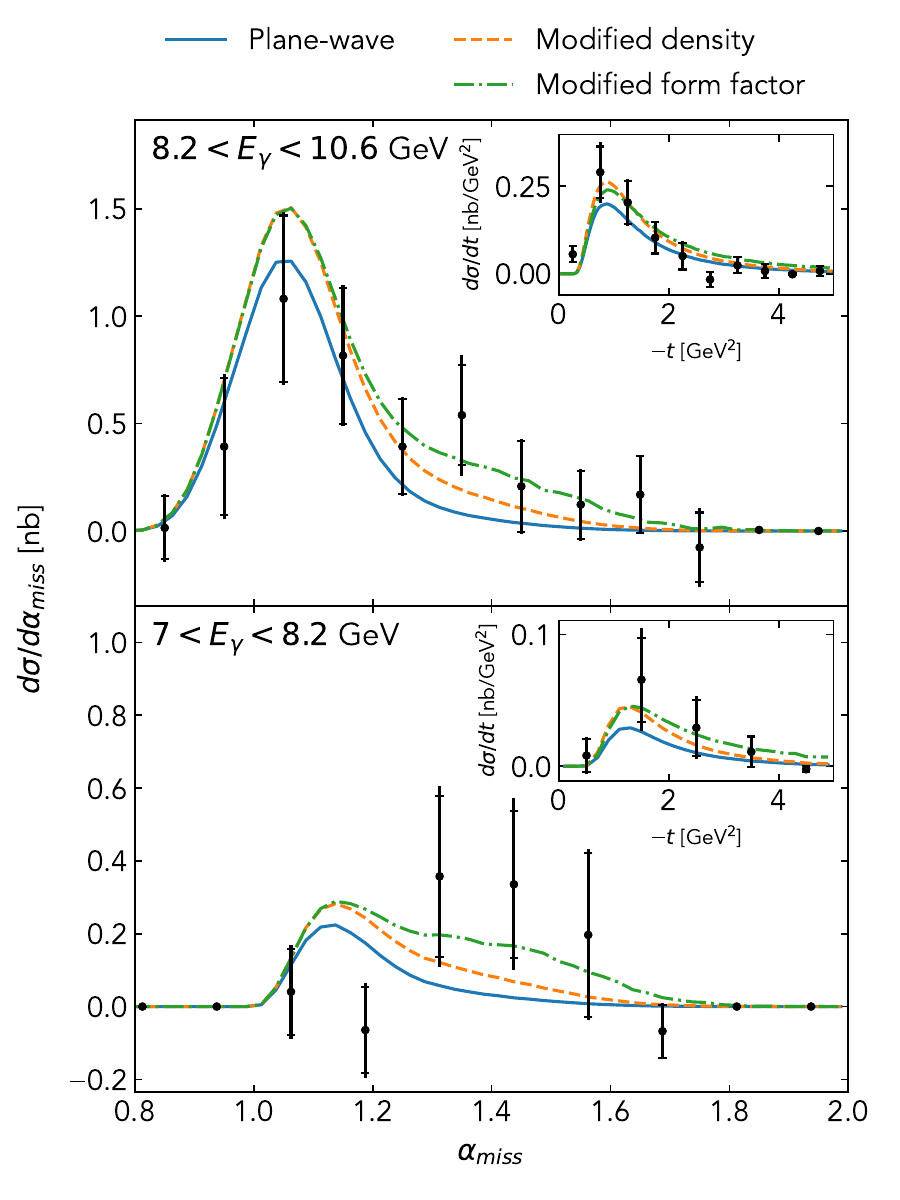}
    \caption{Differential A$(\gamma,J/\psi p)X$ cross section as a function of light-cone momentum fraction $\alpha_\text{miss}$, for $^4$He and $^{12}$C, separated into the above-threshold (top) and below-threshold (bottom) energy regions. Insets shows the differential cross section as a function of momentum transfer $|t|$. Measured data (black points) are compared with plane-wave calculations (blue solid line), as well as calculations assuming a modified proton density (orange dashed) and a modified form factor (green dot-dashed). Data also includes a common 23\% normalization uncertainty (not shown).}
    \label{fig:alphaM}
\end{figure}

In order to further understand the role of nuclear effects present in the reaction, we examine the differential cross section as a function of the struck proton ``missing'' 4-momentum 
\begin{equation}
    p_\text{miss} = p_{e^+} + p_{e^-} + p_{p} - p_\gamma\,.
\end{equation}
Specifically we consider the ``light-cone momentum fraction''  
\begin{equation}
\alpha_\text{miss}=\frac{E_\text{miss}-p_{z,\text{miss}}}{m_A/A}
\end{equation}
of the proton, which gives a measure of the internal nuclear momentum and is well-measured in the GlueX spectrometer.

Figure~\ref{fig:alphaM} shows the differential cross section, extracted separately in the sub- ($7<E_\gamma<8.2$ GeV) and above-threshold  ($8.2<E_\gamma<10.6$ GeV) energy regions.
We note that the limited statistics of the data and the necessity of background-subtraction can result in negative event yields in several kinematic bins, resulting in the extraction of non-physical negative cross sections. 
These negative data points can be interpreted to extract limits on the cross section in a given kinematic bins.

The data are compared with the plane-wave calculations. The inset shows the cross section dependence on momentum transfer $|t|$;
examining the differential cross section in $|t|$ allows for a measure of the gluonic form factor from this data. Assuming a dipole form factor and fitting the plane-wave calculations to the differential cross section between $0<|t|<5$ GeV$^2$, we extract a dipole parameter of $m_s=0.81\pm0.13$ GeV, corresponding to a gluonic radius of $\sqrt{\langle r^2\rangle}=\sqrt{12/m_s^2}=0.85\pm0.14$ fm.
This result is consistent within experimental uncertainties with previous measurements from the proton, such as Refs.~\cite{PhysRevC.108.025201,GravitationalFF} near threshold and Ref.~\cite{PhysRevD.57.1822} at higher energies.
However, this is highly dependent on the limited data at large $|t|$, and a higher-statistics measurement would significantly improve our ability to compare the gluonic radii of free and bound protons.

Considering the $\alpha_\text{miss}$ cross-section dependence we notice an enhancement at large-$\alpha_\text{miss}$ as compared with the PWIA expectation. This shape variation suggests that high-momentum, deeply-bound nucleons could contribute more to the $J/\psi$ cross section than naively expected.
%
%

We examine two possible mechanisms by which the cross section for which deeply-bound protons might be modified. 
The first hypothesis is that the overall scale of the cross section is increased as a function of virtuality, meaning that bound protons couple more strongly to the $J/\psi$.
The second hypothesis is that the $t$-dependence of the cross section is modified, manifesting as a decrease in the slope of the effective gluonic form factor for bound protons.
The former implies an increase in the effective gluon density of the bound proton, while the latter implies a decrease in the effective gluon radius.
In both cases the effect depends on the ``virtuality'' $v=(p_i^2-m_N^2)/m_N^2$ of the bound proton, which quantifies the degree to which the bound proton is off the mass shell and is hypothesized to induce modification of nucleon structure~\cite{Miller_2019}.

The calculated cross sections using either model is shown in Figure~\ref{fig:alphaM}; See Supplemental Material for calculation details.
In both cases the modification to the photon-proton cross section results in an overall increase of the photon-nucleus cross section.
The modification to the form factor predicts a greater increase of the cross section at large $\alpha_\text{miss}$, and is marginally preferred by the data over the plane-wave, but the precision of the current data is insufficient to clearly distinguish between the hypotheses.

Other possible exotic production mechanisms could cause an enhancement in the cross section below threshold.
One possibility is a contribution from hidden-color components of the nucleus, which would enhance coupling via multi-gluon exchanges~\cite{Brodsky_2001}.
Other possible mechanisms include intrinsic charm components of the nucleus, strong binding effects of the $J/\psi$ in cold nuclear matter, or non-local production of $c\bar{c}$ in an extended region of the nucleus.
A more precise measurement of sub-threshold production of $J/\psi$ is called for in order to place more stringent limits on such exotic effects.

In conclusion, we report on the first measurement of $J/\psi$ photoproduction from nuclei at and below the energy threshold of $8.2$ GeV. 
We measure threshold $J/\psi$ cross section for several light nuclei and observe no substantial $A$-dependence.
We extract the energy-dependent cross section for $J/\psi$ production from light nuclei $^4$He and $^{12}$C and compare with plane-wave predictions for near- and sub-threshold production.
In the sub-threshold region, we observe $J/\psi$ production for the first time, and note an excess production relative to theoretical predictions. 
We examine the kinematic distributions of these events and note the sensitivity of sub-threshold $J/\psi$ to modified gluon structure in deeply bound nucleons.

\begin{acknowledgments}

We acknowledge the efforts of the staff of the Accelerator and Physics Divisions at Jefferson Lab that made this experiment possible.
We thank the staff of Jefferson Lab Hall D and many members of the GlueX collaboration for their support of this experiment and their efforts in developing the detector and the analysis tools used here.
This work was supported in part by the U.S. Department of Energy, the U.S-Israel Binational Science Foundation, the Israel Science Foundation, the UK Science and Technology Facilities Council, and the Natural Sciences and Engineering Research Council of Canada.
Jefferson Science Associates operates the Thomas Jefferson National Accelerator Facility for the DOE, Office of Science, Office of Nuclear Physics under contract DE-AC05-06OR23177.
This research also used resources of the National Energy Research Scientific Computing Center (NERSC), a U.S. Department of Energy Office of Science User Facility operated under Contract No. DE-AC02-05CH11231.
This research was supported in part by Lilly Endowment Inc., through its support for the Indiana University Pervasive Technology Institute.

\end{acknowledgments}

\bibliography{references}

\end{document}